\documentclass[aps,twocolumn,groupedaddress]{revtex4}

\usepackage{graphicx}  
\usepackage{dcolumn}   
\usepackage{bm}        
\usepackage{amssymb}   
\usepackage{amsmath}
\usepackage{epsfig}
\usepackage{xspace}
\usepackage{color}

\newcommand{\bea}{\begin{eqnarray}}
\newcommand{\eea}{\end{eqnarray}}

\begin{document}


\title{Static memory materials}
\author{D. Sexty$^{1,2}$}
\author{C. Wetterich$^{3}$}

\affiliation{ $^1$ {\it Department of Physics, Wuppertal University, Gaussstr. 20, D-42119 Wuppertal, Germany}}
\affiliation{ $^2$ {\it J\"ulich Supercomputing Centre, Forschungszentrum J\"ulich, D-52425 J\"ulich, Germany}}
\affiliation{$^3$ {\it Institut f\"ur Theoretische Physik, Universit\"at Heidelberg, Philosophenweg 16, 69120 Heidelberg, Germany}}

\begin{abstract} 

We simulate static memory materials on a two-dimensional lattice. The bulk properties of such materials depend on boundary conditions. 
Considerable information can be stored in various local patterns. We observe local probabilities oscillating with the distance from the boundary. The dependence of the local statistical information on this distance can be described by a linear evolution of classical wave functions, including the superposition principle and classical interference. We speculate that these new phenomena could 
open new algorithmic possibilities analogous to quantum computing. 

\end{abstract}

\maketitle

Memory and information transport are key issues for the exploration of new computational possibilities \cite{1,2,3,4,5,6,7,8}. 
``Static memory materials'' \cite{10} can store information in the equilibrium state of classical statistical systems. This information 
can be imprinted on the boundaries of the material and propagates into the bulk or to another boundary. A key ingredient for the realization 
of a static memory material is a degeneracy of the largest eigenvalue $\lambda_{\rm max}$ of the transfer matrix. The boundary information 
can then be kept within the eigenspace corresponding to the largest eigenvalues, while the part of the information corresponding to smaller 
eigenvalues is lost sufficiently far inside the bulk. More precisely, a static memory material requires more than one eigenvalue $\lambda_i$ 
with $|\lambda_i|=\lambda_{\rm max}$. For complex $\lambda_i$ the local probabilities will show an oscillatory pattern. 

In this note we explore the feasibility of static memory materials by use of numerical Monte-Carlo simulations. Our strategy is a study of 
``imperfect memory materials'' - perhaps closer to the possibilities of actual realization - for which the information is slowly lost as one 
progresses from the boundaries into the bulk. Exact static memory materials are obtained as limiting cases in parameter space. Already for the 
imperfect memory materials we observe well developed new phenomena as oscillating local probabilities, various geometric patterns in the bulk imprinted 
by the boundary conditions, and classical interference. The wave character of the statistical information, organized in probability amplitudes similar to 
quantum mechanics, becomes clearly visible. The analogies of these new structures to quantum mechanics gives hope that static memory materials may offer 
new algorithmic possibilities, similar to quantum computing. 

We first display a few examples of boundary conditions for a two-dimensional imperfect memory material. They demonstrate explicitly the new properties. 
The  formalism for their theoretical description will be sketched subsequently. Our classical statistical equilibrium system is specified
by the probability distribution
\bea\label{XA}
w[s]=Z^{-1}\exp \big(-S[s]\big)
b(s_{in},s_f),
\eea
with partition function
\bea\label{XB}
Z=\int {\cal D} s w[s]
\eea
involving a sum over all spin configurations $\int{\cal D}s$. We discuss an Ising model of spins $s(t,x)=\pm 1 $ on a 2 
dimensional lattice of size $(N_t+1) \times N_x$. The action involves only  diagonal couplings
\bea\label{3}
S= - { \beta \over 2 } 
\sum_{x,t} s(t,x) \Big[s(t+1,x+1) 
+\sigma s(t+1,x-1) \Big]. 
\eea
For $\sigma\neq 1$ it is asymmetric in the two diagonal directions. A technical realization may take a triangular lattice with favoured 
interactions in a given direction. The exact static memory material is realized \cite{10} for $\sigma\to 0, ~\beta\to\infty$. It describes a 
two-dimensional quantum field theory of free massless Weyl fermions \cite{10,25}. The action \eqref{3} involves two independent sub-lattices with ``even'' and ``odd'' lattice sites. Beyond the scope of this note Majorana fermions with left- and 
right-movers can be realized if one interchanges on odd lattice sites $x+1\leftrightarrow x-1$ in eq. \eqref{3}. For $\sigma=1$ one recovers a sum of two two-dimensional Ising models with particular boundary conditions. The parameter $\beta$ can be associated to the inverse temperature in units of the interaction energy. 

The axes for $t$ and $x$ are selected by the specification of boundary conditions at fixed $t$. For simplicity we choose periodic boundary
conditions in $x$, while the boundary conditions for $t_{in}=0$ and $t_f=N_t $ are specified 
by the boundary term
\bea\label{XC}
b(s_{in},s_f)=\bar f_f(s_f)f_{in}(s_{in}),
\eea
with ``initial boundary term'' $f_{in}$ involving only the spins $s_{in}(x)=s(t_{in},x)$,
\bea\label{XD}
f_{in}=\exp\Big(-{\cal L}_{in}(s_{in})\Big),
\eea
and similar for the ``final boundary'', $s_f(x)=s(t_f,x)$,
\bea\label{XE}
\bar f_f=\exp \Big(-{\cal L}_f(s_f)\Big).
\eea

One of the questions asks how the information imprinted on the initial boundary by 
the choice of a specified $f_{in}$ propagates into the bulk, and finally to the 
other boundary where it may be ``read out'' by measuring expectation values $\langle s_f(x)\rangle$. Classical interference is observed if we specify both $f_{in}$ and $\bar f_f$, and read out the information in the bulk. 
For our numerical simulation we employ a Metropolis update. Unless stated otherwise we use $N_t=N_x=32$ and parameters $\beta=4,~\sigma=0.01$. 

We first discuss an ``open final boundary condition'' by choosing $\bar f_f=1$. 
For the ``initial boundary condition'' we start with a fixed configuration of 
initial spins $\bar s_{in}(x)$. This is achieved by 
\bea\label{XF}
{\cal L}_{in}(s_{in})=\lim_{\kappa\to\infty}
\kappa\sum_x\Big(s_{in}(x)-\bar s_{in}(x)\Big)^2.
\eea

\begin{figure}
\begin{center}
\epsfig{file=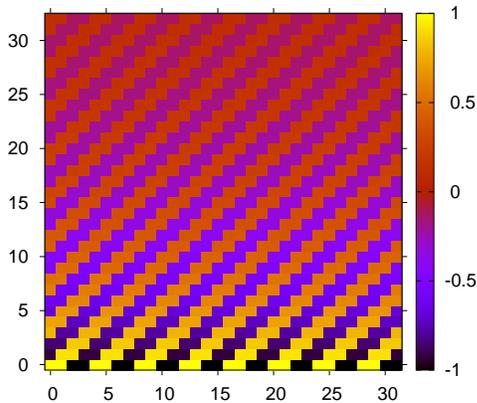, width=8cm}
\caption{ Average spins for fixed initial spins given by eq. \eqref{XG}. The employed set of parameters $N_t=N_x=32,~\beta=4,~\sigma=0.01$ is used in all figures, unless otherwise stated.}
\label{fig-stairs} 
\end{center}
\end{figure}

\begin{figure}
\begin{center}
\epsfig{file=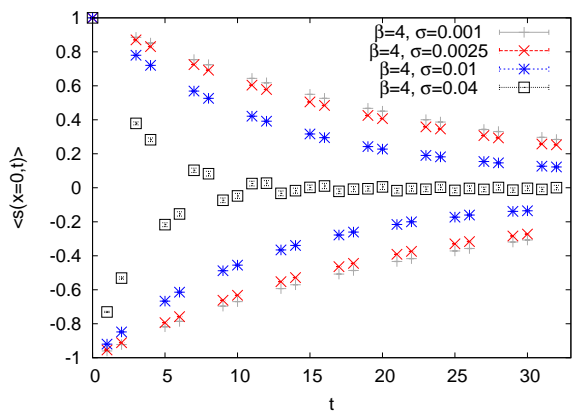, width=8cm}
\caption{ Average spin as a function of $t$ for fixed $x=0$ for $\beta=4$ and four values of $\sigma, ~0,04,~0,01,~0,0025, ~0,001$. The initial boundary conditions are given by eq. (\ref{XG}). }
\label{fig-expdecay} 
\end{center}
\end{figure}

\noindent
The result for an initial condition 
\bea\label{XG}
\bar s_{in}=\left\{ 
\begin{array}{rlll}
1&\text{for}&x=0,1&mod~4\\
-1&\text{for}&x=2,3&mod~4
\end{array}
\right.
\eea
is shown in Fig.~\ref{fig-stairs}. The color code displays the expectation value of each 
individual spin $s(t,x)$ on the lattice. One 
clearly sees how the initial information propagates along the diagonal into the bulk and to the other 
boundary. This system realizes to a good approximation a static memory material.

For any given fixed $\bar x$ the expectation value $s(t,\bar x)$ oscillates with the distance from the boundary $t$, reflecting 
oscillating local probabilities $p(t,x)$ for the spin to be up. We display this oscillation 
in Fig.~\ref{fig-expdecay}. for $\beta=4$ and four values of $\sigma$. As compared to $\sigma=0.01$ an increase of 
$\sigma$ washes out the initial information more rapidly. In contrast, a decrease of 
$\sigma$ preserves information further inside the bulk. It is striking how the asymmetry of the interaction enhances the propagation of information as compared to the Ising model $(\sigma=1)$.
 Around $\sigma=0.001$ a further decrease of $\sigma$ has only a small quantitative influence. In this region the damping of 
the initial information is mainly due to the finite value of $\beta$. Increasing $\beta$ for 
$\sigma=0.001$ results in an even slower loss of the initial information. All this is in 
accordance with the observation that for $\beta\to\infty,\sigma=0$ 
our system is an exact static memory material without any loss of information. 


Our second example associates uncorrelated weight factors to the initial spins
\bea\label{XH}
f_{in}(s_{in})=\prod_x 
\Big[\bar p_+(x)h_+\big(s_{in}(x)\big)+\bar p_-(x)h_-\big(s_{in}(x)\big)\Big],
\eea
with 
\bea\label{XI}
h_\pm\big(s(x)\big)=\frac12\Big(1\pm s(x)\Big),\quad 
\bar p_\pm(x)=\frac12\Big(1\pm\bar s(x)\Big),
\eea
and $0\leq\bar p_\pm(x)\leq 1$. For open final boundary conditions $(\bar f_f=1)$ the relative probabilities to find $s(x)$ up, 
as compared to down, is given by $\bar p_+(x)/\bar p_-(x)$. One finds for open final boundary conditions $\langle s_{in}(x)\rangle=\bar s(x)$. ``Wave boundary conditions'' are specified by
\bea\label{XL}
\bar s(x)=\sin 
\left(\frac{2\pi m x}{N_x}\right), ~m \in \mathcal{Z}.
\eea

\begin{figure}
\begin{center}
\epsfig{file=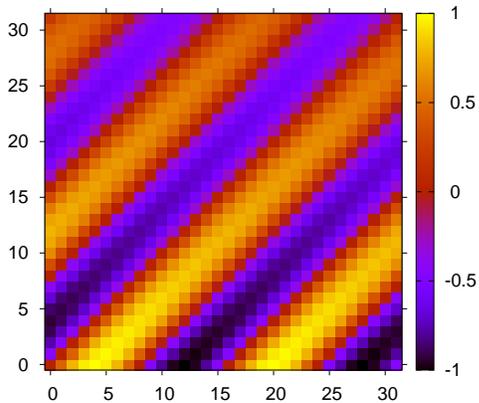, width=8cm}
\caption{ Average spins for an initial wave boundary condition with $m=2$ and open final boundary condition.}
\label{fig-wave} 
\end{center}
\end{figure}

\noindent
They imprint on the system oscillating local probabilities, with periods in $t$ 
given by $N_x/m$. 
In Fig.~\ref{fig-wave}. we show the result of an initial wave boundary condition with $m=2$. By comparison with Fig. \ref{fig-stairs} one observes that the memory of larger structures is conserved further inside the bulk \cite{10}. Indeed, an initial wave boundary condition with $m=4$ shows a loss of information more similar to Fig. \ref{fig-stairs}. 
Together with Fig.~\ref{fig-stairs}, it is obvious that very different local patterns can be realized by 
the memory material by choosing appropriate boundary conditions. 

\begin{figure}
\begin{center}
\epsfig{file=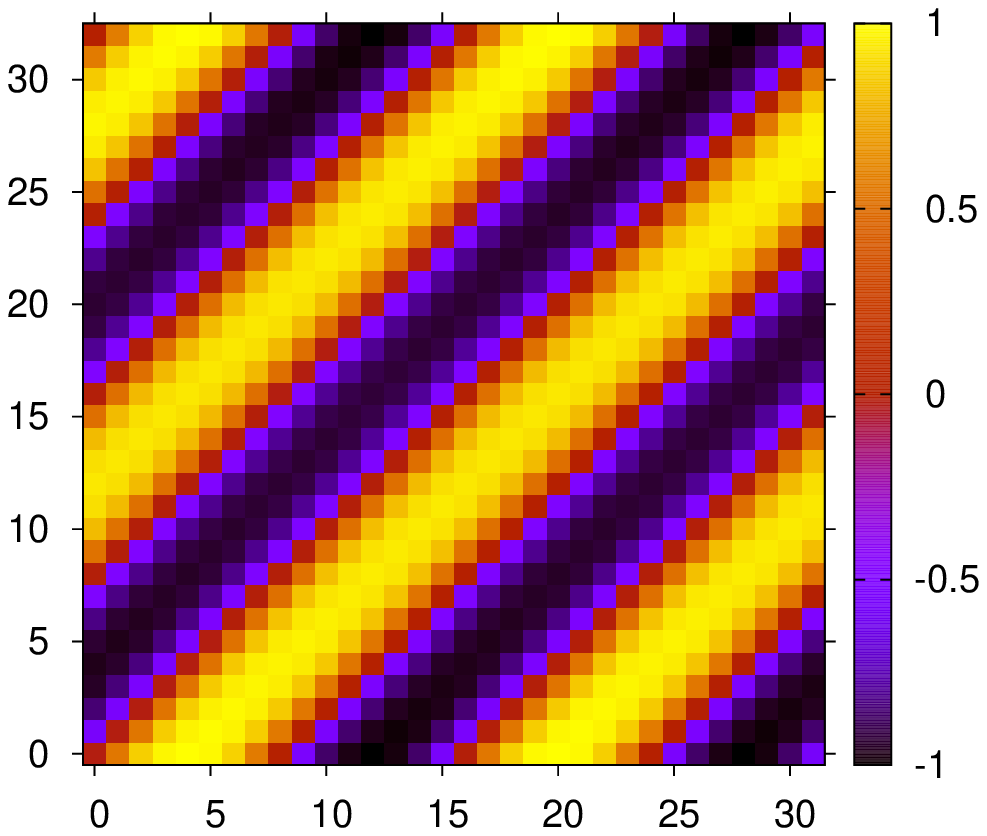, width=8cm}
\epsfig{file=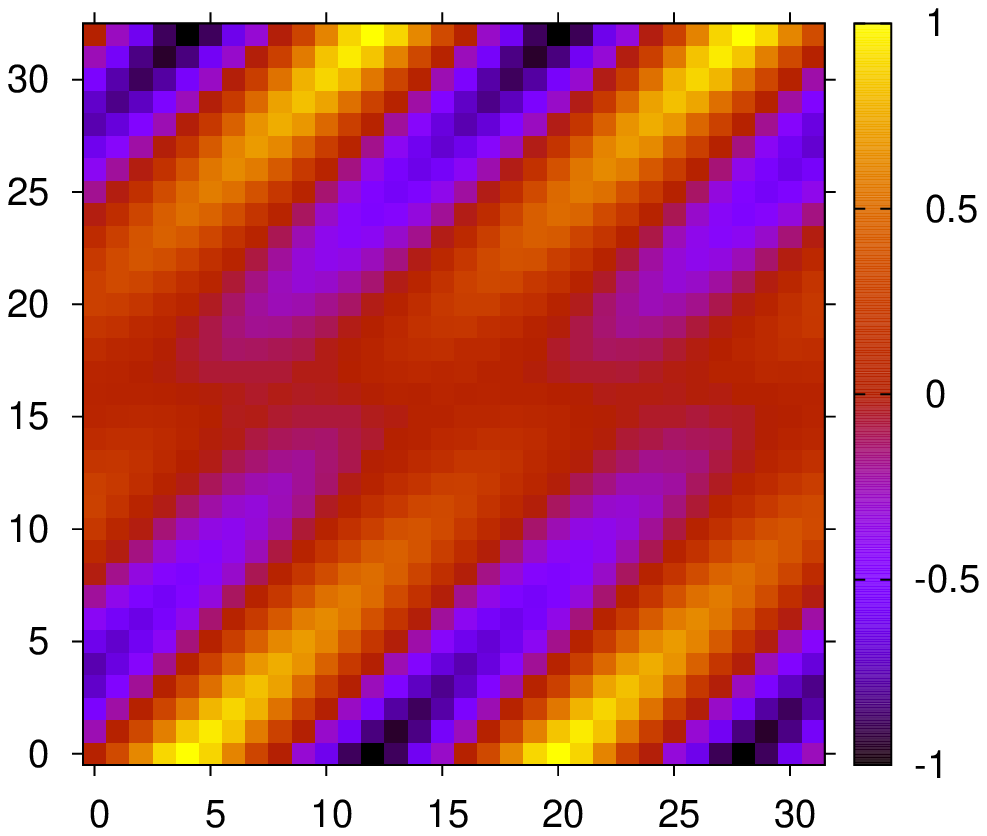, width=8cm}
\caption{ Average spins for initial wave boundary condition with $m=2$, and final wave boundary condition with $m=2$ (top) and $m=-2$ (bottom).}
\label{fig-1part-open} 
\end{center}
\end{figure}

\begin{figure}
\begin{center}
\epsfig{file=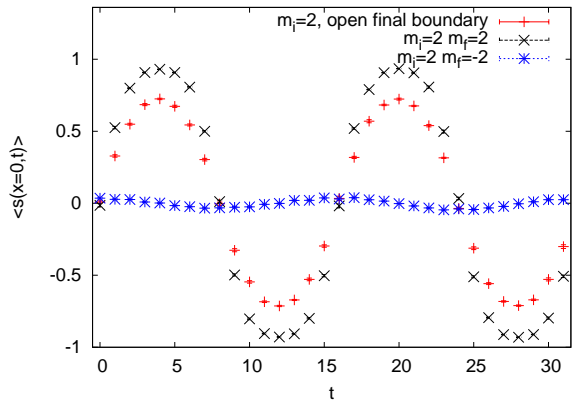, width=8cm}
\caption{ Average spins at $t=16=(t_i+t_f)/2$ as a function of the $x$ coordinate. 
The initial boundary condition is a wave given by eqs. (\ref{XH}-\ref{XL}) with $m=2$. We display results using 
three different 
final boundary conditions, namely open or waves with $m=2$ and $m=-2$. }
\label{middle-slice} 
\end{center}
\end{figure}

We next impose boundary conditions both at the initial and final boundary. The ``readout'' of information 
has now to proceed by measuring expectation values $\langle s(\bar t,x)\rangle$ for some $\bar t$ inside 
the bulk. In Fig. \ref{fig-1part-open} we display the result for the same initial boundary condition as for Fig. \ref{fig-wave} (wave boundary condition with $m=2$), but now with final boundary conditions given by the waves $(9-11)$ with $m=2$ and $m=-2$. Comparison with Fig. \ref{fig-wave} demonstrates the importance of the final boundary condition. Classical interference becomes visible by comparison of the final boundary conditions. For $m=2$ the interference is positive and the preservation of information inside the bulk is enhanced. In contrast, the negative interference for $m=-2$ reduces the available information in the center of the bulk. We plot the average spins in the middle of the bulk in Fig. \ref{middle-slice}, with three boundary conditions corresponding to Figs. \ref{fig-wave} and \ref{fig-1part-open}. This demonstrates quantitatively the role of classical interference. 

Ising spins are associated to bits or fermionic particles, with a particle present for $s=1$ and absent for $s=-1$. The local spins directly correspond to local occupation numbers $n(t,x)=\big(s(t,x)+1\big)/2$. We may define a particle number $N_p(t)$ by the total number of spins up at $t$,
\bea\label{YA}
N_p(t)=\frac12\sum_x\Big(s(t,x)+1\Big).
\eea



\noindent
One particle states correspond to configurations where only one spin is up, while all others are down. 
For an initial state with a single particle at $y$ the initial boundary term $f_{in}$ is proportional to
\bea\label{YB}
h_1(y)=n(y)\prod_{x\neq y}\Big(1-n(x)\Big).
\eea
An initial one particle state is given by the boundary condition
\bea\label{YD}
f_{in}=\sum_yq_1(t_{in},y)h_1(y)~,~q_1(t_{in},y)\geq 0.
\eea
In contrast to the boundary condition \eqref{XH}, the spins of a one particle initial state are highly correlated. 
Whenever a spin at one position is up, all other spins are down. We investigate ``wave-packet boundary conditions'' somewhat analogous to the description of particles in quantum mechanics. 
\bea\label{XAA}
q_1(x)=\exp\left( {-(x-x_0)^2 \over 2 \Delta^2 } \right). 
\eea

\begin{figure}
\begin{center}
  \epsfig{file=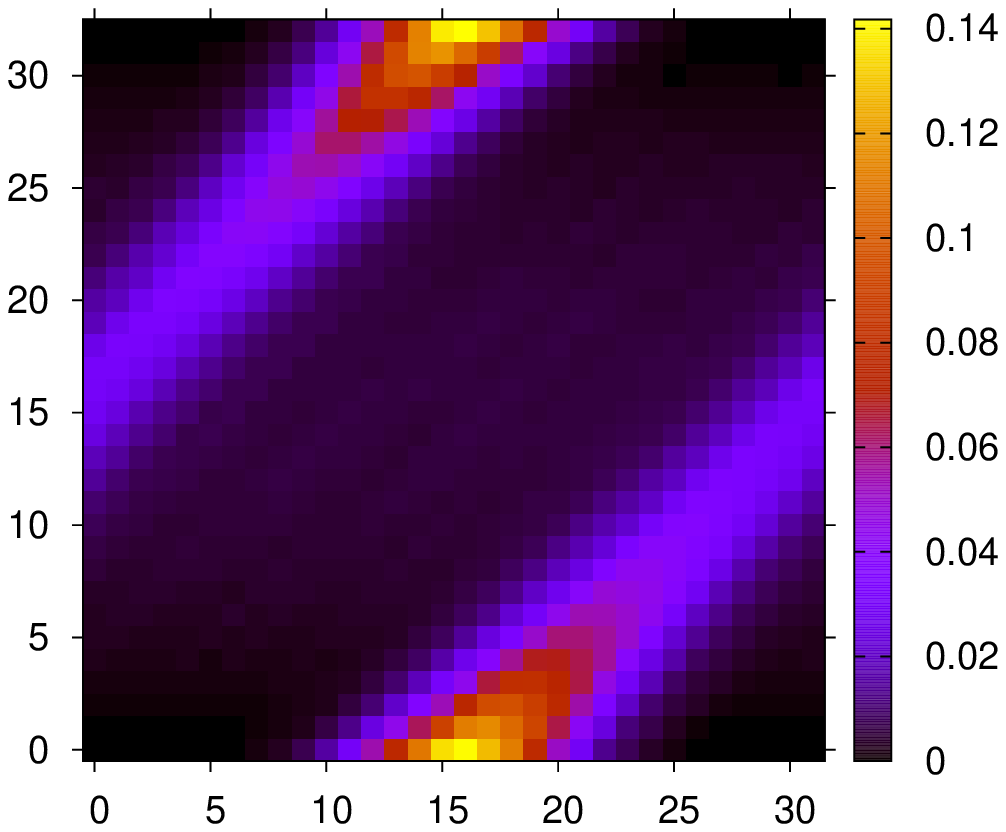, width=8cm}
  \epsfig{file=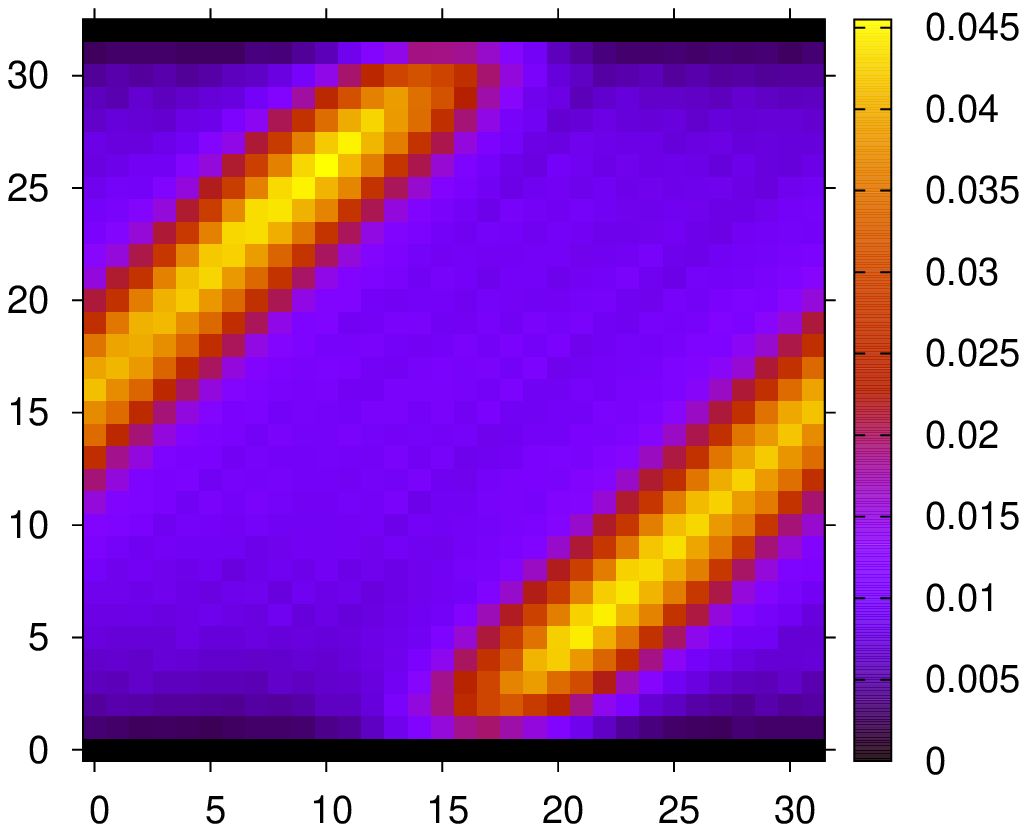, width=8cm}
\caption{ The observable $n_1, n_2$ on a $N_x=32,N_t=32$ lattice. The parameters are $\beta=4$, $\sigma=0.01, x_0=16, \Delta=3 $. The initial and final boundary conditions are wave-packet one-particle states. 
}
\label{plot_gauss} 
\end{center}
\end{figure}

\begin{figure}
\begin{center}
  \epsfig{file=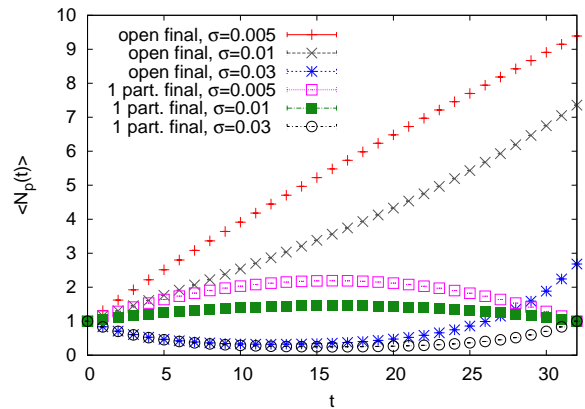, width=8cm}
  \caption{ The average particle number as a function of $t$ for various wave-packet boundary conditions and $\sigma$ parameters. The initial boundary condition is a one-particle state given by eqs. (\ref{YD}-\ref{XAA}). The final boundary condition is either open or also given by a one particle wave-packet, as indicated.}
\label{plot_gaussnp} 
\end{center}
\end{figure}

A useful observable for one particle states is the one-particle local occupation number 
\bea\label{A3}
n_1(t,x)=
\left\{ 
\begin{array}{cll}
\big(s(t,x)+1\big)/2&\text{if}&N_p(t)=1\\
0&\text{for}&N_p(t)\neq 1.
\end{array}
\right.
\eea
It differs from zero only if precisely one particle is present at $t$, and measures the probability to find this particle at $x$. A similar quantity $n_2$ for two particle states replaces $N_p=2$ in eq. \eqref{A3}. In Fig. \ref{plot_gauss} we plot $n_1$ and $n_2$ for Gaussian initial and final boundary conditions. This can be interpreted as a single particle decaying into two collinearly moving particles. For the sum $n_1+n_2$ the information loss is rather moderate.
We show the average particle number $ \langle N_p(t) \rangle $ as a function of the coordinate $t$ for various boundary conditions in Fig.~\ref{plot_gaussnp}. 
The non-conservation of the particle number is an effect of the finite $\beta$ as well as the interactions given  by $\sigma>0$.




For an analytical description we follow the quantum formalism of refs. \cite{10,CWQF}. The transport of information from a hypersurface 
at $t$ to a neighboring one at $t+1$ can be described by the transfer matrix. We restrict the discussion here to the part of the transfer matrix which is projected on one-particle states. We note that particle number is only approximately conserved. For finite $\beta$ a particle can split into two or even more particles. Since these particles move collinearly many aspects are similar to the evolution of one particle states for which the influence of particle number non-conservation is neglected. For our model \eqref{3} the projected transfer matrix reads 
\bea\label{C1}
\bar S(x,y)&=&\bar {\cal N}\Big[(1-e^{-2\beta})\delta(x,y+1)\\
&&+e^{-2\beta}(e^{2\beta\sigma}-1)\delta(x,y-1)+e^{-2\beta}\Big],\nonumber
\eea
with $\bar {\cal N}=\exp \left\{ \frac\beta2\Big[N_x+\sigma(N_x-4)\Big]\right\}$. Here we use for every $t$ the ``one-particle location basis'' with 
basis functions $h_1(x)$ defined by eq. \eqref{YB}. The step evolution operator $S$ \cite{10} renormalizes the transfer matrix multiplicatively 
such that the absolute value of the largest eigenvalue $\lambda_{\rm max}$ is set to one. Thus the multiplicative factor $\bar {\cal N}$, which corresponds 
to an additive constant in eq. \eqref{3}, does not matter. In the limit $\beta\to\infty$ one finds for all $\sigma<1$ the simple 
expression $S(x,y)=\delta(x,y+1)$. The eigenvalues of $S$ obey then $\lambda^{N_x}=1$, and therefore $\lambda_M=\exp (2\pi i M/N_x)$. This system keeps 
complete memory of the boundary conditions. For large $\beta$ the step evolution operator $S$ replaces in eq. \eqref{C1} the factor $\bar{\cal N}$ by 
${\cal N}=\left[1+e^{2\beta(\sigma-1)}+(N_x-2)e^{-2\beta})\right]^{-1}$. 

Similar to eq. \eqref{YD} we may define for each $t$ a classical one-particle wave function $q_1(t,x)$. The evolution with $t$ is described by multiplication 
with the step evolution operator \cite{10},
\bea\label{C2}
q_1(t+1,x)=\sum_yS(x,y)q_1(t,y),
\eea
with initial condition at $t_{in}$ given by $q_1(t_{in},y)$ according to eq. \eqref{YD}. This constitutes a linear evolution law, with 
\bea\label{C3}
q_1(t_{in}+a,x)=\sum_y
S^a(x,y) q_1(t_{in},y).
\eea
Similarly, one has for the conjugate wave function
\bea\label{C4}
\bar q_1(t_f-b,x)=\sum_y\bar q_1(t_f,y)
S^b(y,x),
\eea
with $\bar f_f=\sum_y\bar q_1(t_f,y)h_1(y)$. The local probability $p_1(t,x)$ to find at $t$ a particle (or spin up) at the position $x$ 
is a bilinear in the classical wave functions
\bea\label{C5}
p_1(t,x)=\frac{1}{N_1}\bar q_1(t,x)q_1(t,x),
\eea
with $N_1=\sum\limits_x\bar q_1(t,x)q_1(t,x)$. The linear evolution \eqref{C2} of $q_1$ entails the superposition principle for solutions, 
and similar for $\bar q_1$. Together with the bilinear expression \eqref{C5} this leads to interference effects well known from quantum mechanics. The particle propagation for the present Ising model differs from quantum particles, due to the presence of two different wave functions $q$ and $\bar q$ that are both positive and real, as compared to the complex wave function for quantum mechanics. This limits the observable interference patterns. 

For sufficiently smooth wave functions we may define a continuum limit
\bea\label{C6}
\partial_tq_1=\frac12 \Big[q_1(t+1)-q_1(t-1)\Big]=W q_1(t),
\eea
with 
\bea\label{C7}
W=\frac12(S-S^{-1}).
\eea
Using for large $\beta$ the leading terms in eq. \eqref{C1} yields
\bea\label{C8}
\partial_tq_1(x)&=&-
\left[1-\frac{e^{-2\beta}}{2}(e^{2\beta\sigma}-1)\right]\partial_x
q_1(x)+D,\nonumber\\
D&=&\frac{e^{-2\beta}}{2}
\Big\{(e^{2\beta\sigma}+1)\sum_yq_1(y)\nonumber\\
&&-(3e^{2\beta\sigma}+2N_x-3)q_1(x)\Big\}.
\eea
The first term corresponds to the antisymmetric part of $W$ and accounts for the propagation of the particle to the right. 
For $\beta\to\infty$ this is the only term. The term $D$ arises from the symmetric part of $W$. It accounts for the loss of boundary information. 
For $2\beta\sigma\ll 1$ one has $D\approx e^{-2\beta}\sum_y\Big[q_1(y)-q_1(x)\Big]$. The conjugate wave function obeys
\bea\label{C9}
\partial_t\bar q_1=-W^T\bar q_1,
\eea
such that the term $D$ changes sign as compared to eq. \eqref{C8}. 

For large enough $\beta$ all terms involving factors $e^{-2\beta}$ can be neglected. Then both $q_1$ and $\bar q_1$ obey the same evolution equation
\bea\label{C10}
(\partial_t+\partial_x)q_1=0,~(\partial_t+\partial_x)\bar q_1=0.
\eea
In this limit the particle number $N_p$ is conserved. For large enough $\beta$ the influence of 
configurations with $N_p\neq 1$ can therefore be neglected for all $t$ if the initial and final 
boundary terms are given by one-particle states. For $\beta\to\infty$ and boundary conditions 
$\bar q(t_f)=q(t_f)$ one has $\bar q(t)=q(t)$ for all $t$. The system \eqref{C10} describes the unitary 
evolution of a quantum system. (The more familiar complex formulation can be found in ref. \cite{10}, and 
the equivalence with a two-dimensional fermionic quantum field theory is established in ref. \cite{25}.) For 
our choice of parameters the approximation \eqref{C10} is already rather accurate, if no distinction between a single particle and a collinear multiparticle state is made, cf. Fig. \ref{plot_gauss}.

The important role of wave functions and the observed interference effects lead to the speculation that static memory materials can be used for the 
implementation of algorithms similar to quantum computing. We observe that for asymmetric couplings $(\sigma<1)$ the technical 
realization of static memory materials seems feasible if large enough $\beta$ can be realized. Static memory materials can be realized 
under a wide range of circumstances - they are not restricted to two dimensions. We hope that in the future they can find technical applications.

\acknowledgments \noindent

D. Sexty is funded by the 
Heisenberg programme of the DFG (SE 2466/1-1). C.~Wetterich acknowledges the support from DFG Collaborative Research center SFB 1225 (ISOQUANT).

\bibliography{mybib}
  
\end{document}